\documentclass[aps,pre,twocolumn]{revtex4}
\usepackage{graphicx}
\usepackage{xcolor}
\usepackage{amsmath}
\usepackage [utf8]{inputenc}
\usepackage[capitalize]{cleveref}
\def\beq{\begin{equation}}
\def\eeq{\end{equation}}
\def\bea{\begin{eqnarray}}
\def\eea{\end{eqnarray}}

\begin{document}
\makeatletter
\title{Coherence of velocity fluctuations in a turbulent round jet} 

\author{T. Singla}
\affiliation{Laboratoire de Physique de l'\'Ecole Normale Sup\'erieure, CNRS, PSL Research University, Sorbonne Universit\'e,
Universit\'e de Paris, F-75005 Paris, France} 

\author{M. Caelen}
\affiliation{Laboratoire de Physique de l'\'Ecole Normale Sup\'erieure, CNRS, PSL Research University, Sorbonne Universit\'e,
Universit\'e de Paris, F-75005 Paris, France} 

\author{F. P\'etr\'elis}
\affiliation{Laboratoire de Physique de l'\'Ecole Normale Sup\'erieure, CNRS, PSL Research University, Sorbonne Universit\'e,
Universit\'e de Paris, F-75005 Paris, France} 

\author{S. Fauve}
\affiliation{Laboratoire de Physique de l'\'Ecole Normale Sup\'erieure, CNRS, PSL Research University, Sorbonne Universit\'e,
Universit\'e de Paris, F-75005 Paris, France} 

\date{\today}

\begin{abstract}
Coherence between the velocity fluctuations measured at two points is investigated in a turbulent jet flow. Coherence is calculated in different regions of the flow by changing the separation between the points, the distance between the points from the exit of the jet and from its central axis, and the injection velocity of the jet. It is observed that coherence scales with the system parameters, the functional form of coherence and the corresponding scaling laws change depending on in which region coherence is measured, and that coherence is self similar in both the regions.
\end{abstract}

\maketitle
\section{Introduction}

The Richardson-Kolmogorov theory of turbulence \cite{Kolmogorov} explains how energy is transferred from the integral length scale eddies to the smaller ones until it starts to dissipate in the form of heat at smaller Kolmogorov scales. In the context of fully developed turbulent flow, the spectrum of length scales spanning from the integral scales to the Kolmogorov length scales constitutes what is commonly referred to as the inertial range. The statistical properties of the turbulent flow within the inertial range have been thoroughly studied, however, for scales larger than the integral lengths, understanding of these properties is still poor. Recently, there have been a few works in which turbulent flows have been experimentally investigated where large scale dynamics of the turbulent flows have been studied. Coherence is calculated between the velocity fluctuations at two points and large scale dynamics was explained \cite{Gaurav}. In another study, it has been shown that the large scale dynamics remain in statistical equilibrium when energy is injected at scales smaller than integral length scales \cite{Gorce}.

A turbulent jet refers to the type of fluid flow in which a high velocity fluid is injected into a surrounding fluid with low velocity. The shear forces due to mixing of these fluids cause the formation of turbulent eddies provided that the Reynolds number of the injected flow is sufficiently high. Different studies carried out to study turbulent jets involve understanding their spectral analysis \cite{Gibson,Schmidt}, the dependence of flow dynamics on the shape of the jet nozzle \cite{Mi}, and the effects of intermittency and shear on the scaling exponents of turbulent jet \cite{Mi1}. The flow of a turbulent jet can be characterized by several parameters, namely, jet diameter at flow injection ($D$), axial distance ($x$: along the direction of flow), radial distance ($r$: perpendicular to $x$), and half width ($r_{1/2}$: distance at which the mean jet velocity is half of the mean velocity at the central axis).

It is known, that a turbulent flow is self-similar within the inertial range and the energy transfer processes and statistical properties within this range exhibit a universal behavior that is independent of the particular flow configuration or initial conditions. A turbulent flow is called self-similar when some of its statistical properties can be determined by specific combinations of control parameters, rather than each parameter individually. In other words, properties at two different points of a self-similar flow can be related by a scaling factor given by the combination of parameters. This makes the analysis of turbulent flow easier, especially in experimental studies as they rely on long duration measurements for the convergence of the properties. The concept of self-similarity in turbulence is known since 1938 when K\'arman and Howarth \cite{Karman} studied correlation functions in decaying grid turbulence. In subsequent works, George \cite{George} and Speziale and Bernard \cite{Speziale} showed that self-simillarity of the flow indicates a power law decay of the mean turbulent energy. The universal nature of self-similarity in the flow has also been explored. It is reported that self-similarity in turbulent flows can be either partial i.e. not valid throughout the entire range of scales \cite{Meldi} or complete \cite{Tang}. Gonzalez and Fall \cite{Gonzalez} showed that complete self-similarity requires $\langle k\rangle\propto t$, which implies that the Reynolds number remains constant during the evolution of the flow. It is known that turbulent jets meet this criterion and hence complete self-similarity has been reported in the flows generated by them \cite{Thiesset,Gauding}.

Coherence ($C$) is a quantity that is useful to understand the relations between the spatio-temporal scales of turbulent flow. It is usually calculated between velocity fluctuations ($v_1$ and $v_2$) acquired at two points separated by a distance $r$ and it is a function of frequencies of the eddies present in the flow. It can be calculated with the following expression:
\begin{eqnarray}
C_{v_iv_j}(r,f)&=&\frac{|G_{v_iv_j}(r,f)|^2}{G_{v_iv_i}(r,f)G_{v_jv_j}(r,f)}\\
G_{v_iv_j}(f)&=&\int_{-\infty}^{\infty} \langle v_i(t)v_j(t+\tau)\rangle e^{-i2\pi f\tau}d\tau
\end{eqnarray}
 where, $i, j = 1,2$ and $G_{v_iv_j}(f)$ represents the auto/cross power spectral density of the signals depending on $i$ and $j$ being equal/different. $C$ takes a value of $1$ if two signals at certain frequencies ($f$) are perfectly correlated and $0$ if there is no correlation at all. 

Coherence is an important quantity for the estimation of power load fluctuations in wind turbine farms \cite{Vermeer,Sorensen} or to evaluate large scale constraints on bridges \cite{Cheynet} and buildings \cite{Emil}. The coherence function is, however, still poorly documented, with measurements only in the context of the turbulent atmospheric boundary layer and turbulent wakes. Various empirical models have been  proposed to describe the coherence between velocity fluctuation in a turbulent flow. Davenport \cite{Davenport} propose $C(r,f) = e^{-arf/U_{rms}}$, where $U_{rms}$ is the root-mean-square velocity of the flow at the location of measurements. Thresher et al. \cite{Thresher,Saranyasoontorn} proposed another model involving the length scales of the flow such that $C(r,f) = (e^{-a\sqrt{(rf/U_{rms})^2 + (br/L_c)^2}})^2$. In a theoretical work, Tobin and Chamorro \cite{Tobin} proposed another model for coherence in a high mean velocity turbulent flow. In this case $C(r,f) \propto e^{-ar^2f^2/U_{rms}^2}$.

The aim of this paper is to investigate coherence in the flow generated by a turbulent round jet. Coherence is calculated between the velocity fluctuations at two points separated by certain distance. Velocities are measured separately in two different regions defined by the central axis line and the line that follows half-width along the central axis. From the measurements of coherence a mathematical form is first obtained for it in both of the regions and then scaling laws were identified to establish its self-similar nature. It is also observed how the nature of coherence changes, notably its value at zero frequency and its decay, from one region to the other. The next section of this paper shows the experimental setup that is used to measure the velocities at two points. Two regions that are explored to study the coherence are also shown on this setup. This section is followed by results for both the regions where the functional form of coherence and different scaling laws demonstrating self-similarity are derived from the velocity measurements. Finally, the paper ends with the conclusions section where important observations from different results are presented.

\begin{figure}[ht!]
\includegraphics[width=6cm,height=4cm]{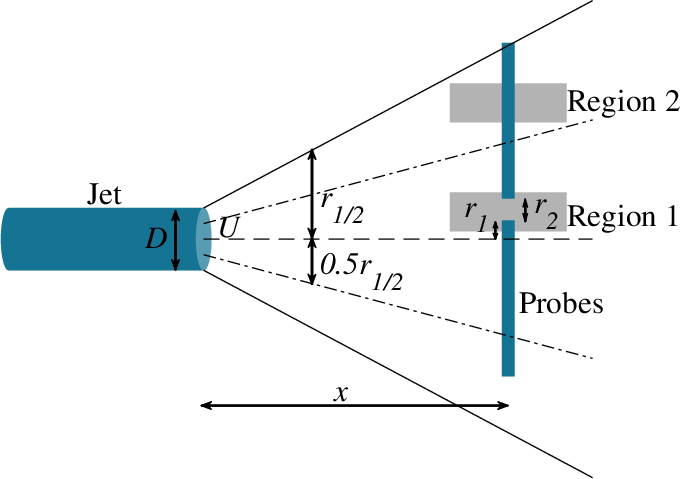}
\caption{Experimental setup demonstrating the tube of diameter $D$ that was used to create a turbulent jet flow. Measurements were performed at the central plane of the flow using hot-wire anemometers (represented as probes). $U$ represents the velocity of the flow at the nozzle of the tube, $r_{1/2}$ is the line of half-width, $r_1$ is the distance of the probes from the central axis, $r_2$ is the probe separation, and $x$ is the distance of the probes from the nozzle of the jet. Coherence is calculated and analysed separately in Region 1 and Region 2.}
\label{setup}
\end{figure}

\section{Experimental setup}

Fig. \ref{setup} shows the experimental setup of the turbulent jet. It consists of a tube of length 10cm and inner diameter $D = 1$cm. Compressed air was introduced into the surrounding area with a velocity ($U$) to create flow. The range of air ejection velocity was kept between $U = 15$m/s to $30$m/s for different measurements. This sets the Reynolds number of the flow between $R = UD/\nu$ between 15000 and 30000, which is sufficiently high to create a turbulent flow. Two DANTEC 55P16 hot wire probes were used in front of the jet to measure the velocities, and a DANTEC StreamWare Pro system coupled to a NI DAQ card was used to convert the velocity into voltage signals. The probes were calibrated before taking measurements in order to convert the voltage signals to the velocity of the flow. The positions of the probes were controlled by changing their distance from the jet ($x$), 1st probe to central axis distance ($r_1$), and probes separation ($r_2$). For a fixed value of $x$, $r_{1/2}$ on Fig. \ref{setup} indicates the radial distance from the central axis at which mean velocity of the flow is half of the mean velocity at the central axis. As reported by \cite{Samie} et al. another quantity that is important in the context of coherence of velocity fluctuations in a turbulent jet is $0.5r_{1/2}$. Different measurements are taken by keeping both the probes either in region 1 lying between the central axis and the $0.5r_{1/2}$ line, or in region 2 (between $0.5r_{1/2}$ and $r_{1/2}$ line). The probes were fixed such that their resistor filaments remained perpendicular to the flow, allowing us to measure the component of the velocities in the direction of the flow.

\begin{figure}[ht!]
\includegraphics[width=4cm,height=4cm]{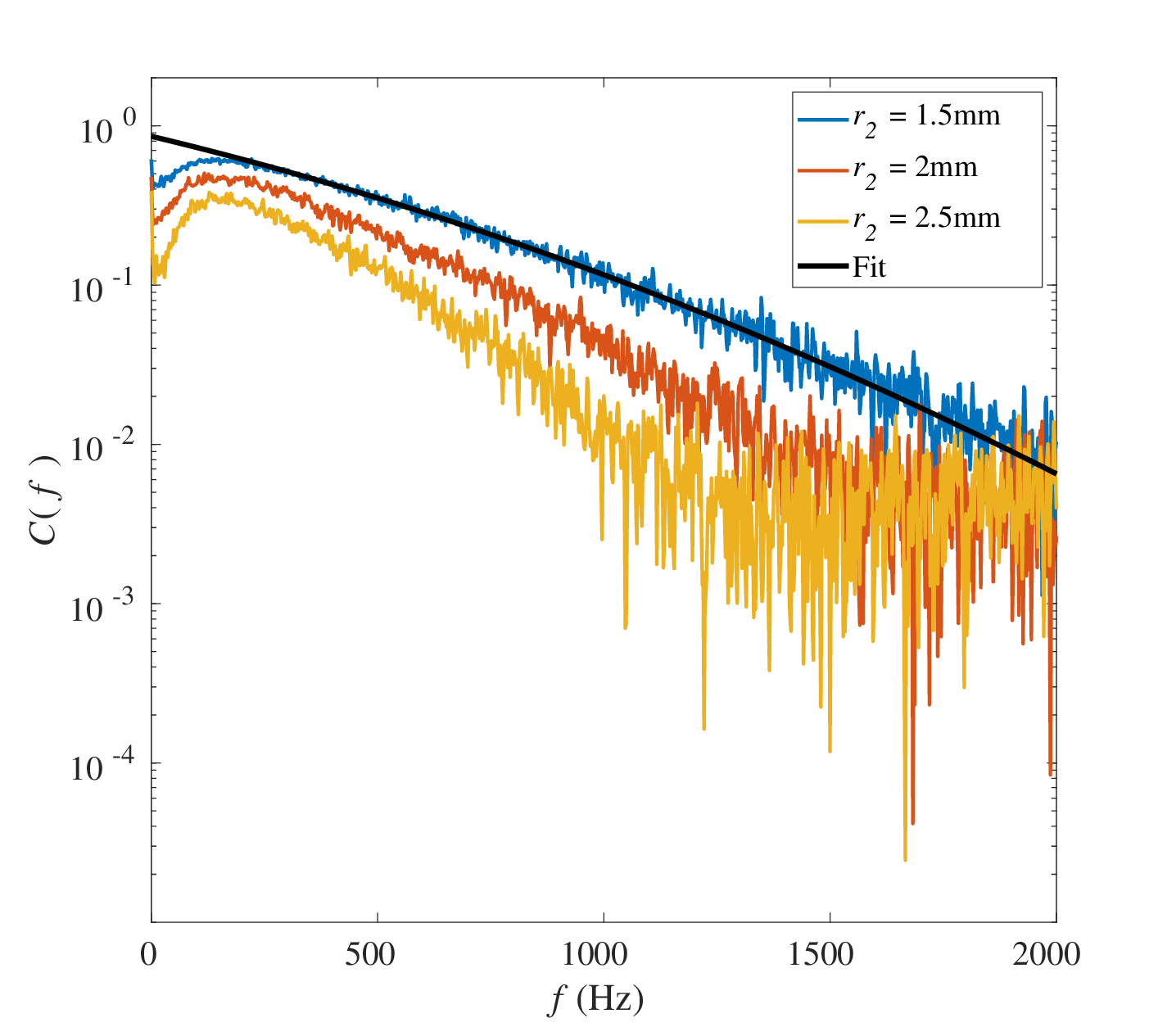}
\includegraphics[width=4cm,height=4cm]{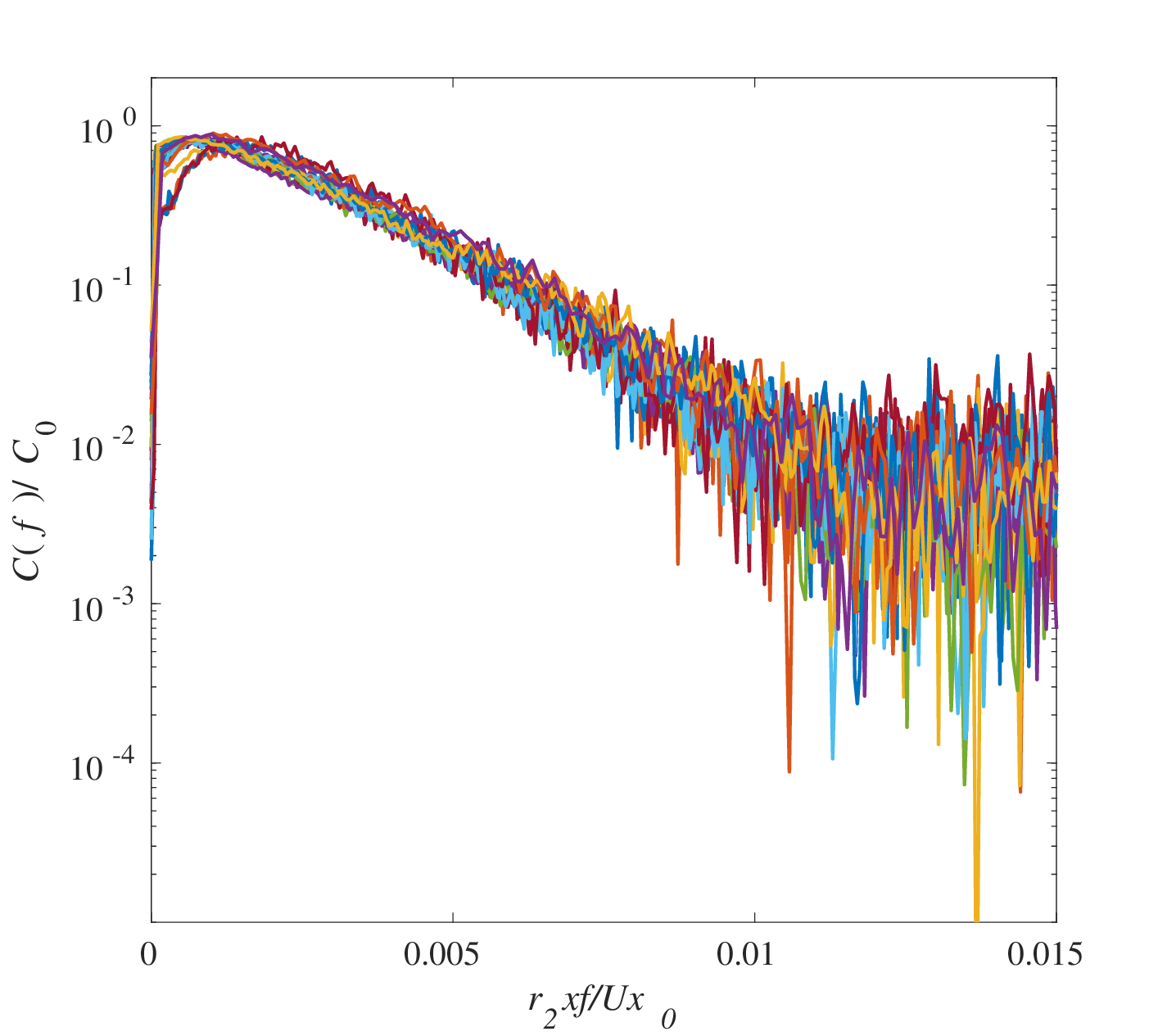}
\caption{(Left panel) Coherence curves in Region 1 of the turbulent jet as a function of $r_2$. Black curve represents a fit on one of the coherence curves and its functional form is given by Eq. \ref{eq3}. (Right panel) Re-scaled coherence curves for several parameters of the turbulent jet.} 
\label{fig2}
\end{figure}

\section{Results}

\begin{figure*}[ht!]
\includegraphics[width=5cm,height=5cm]{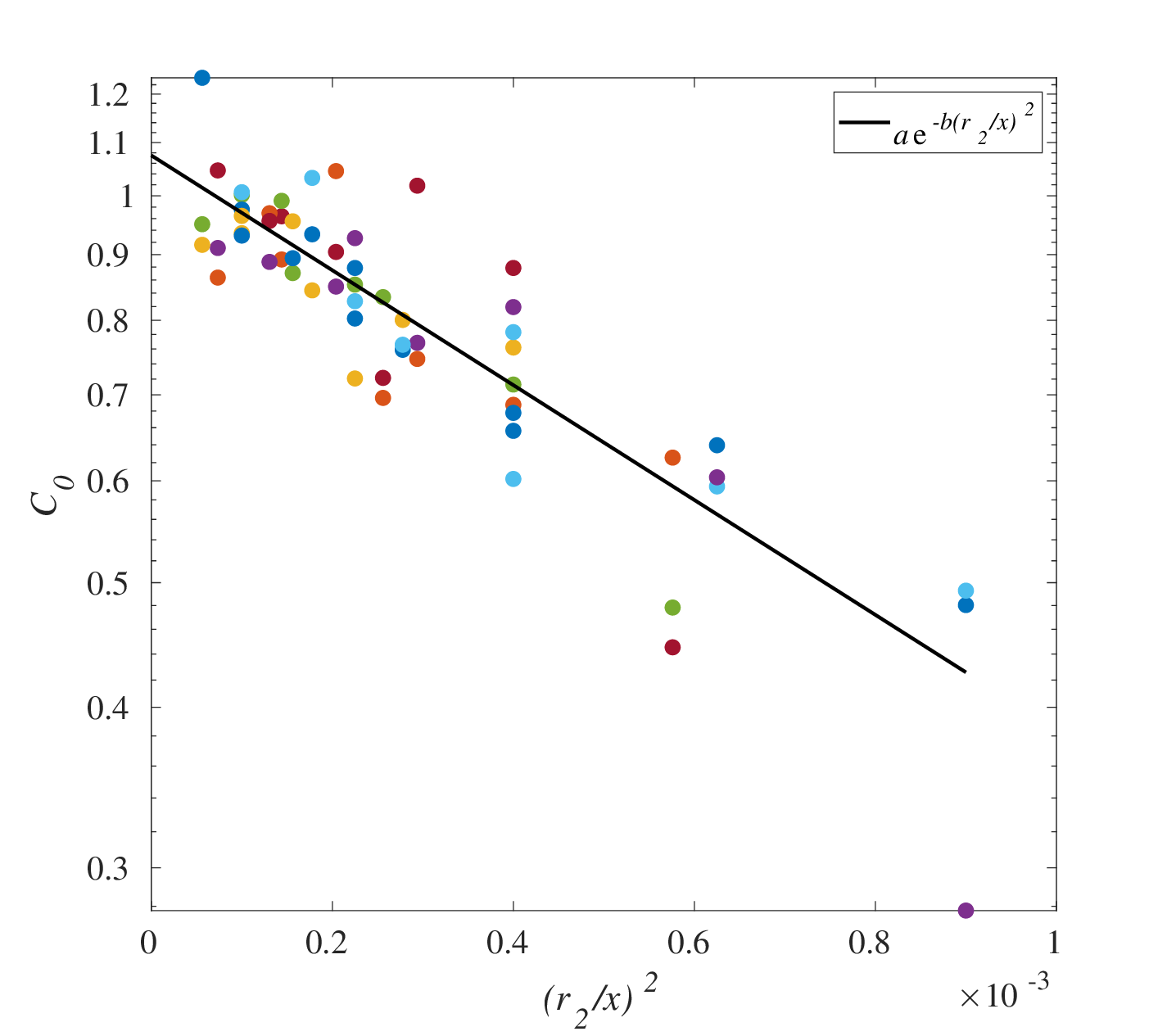}
\includegraphics[width=5cm,height=5cm]{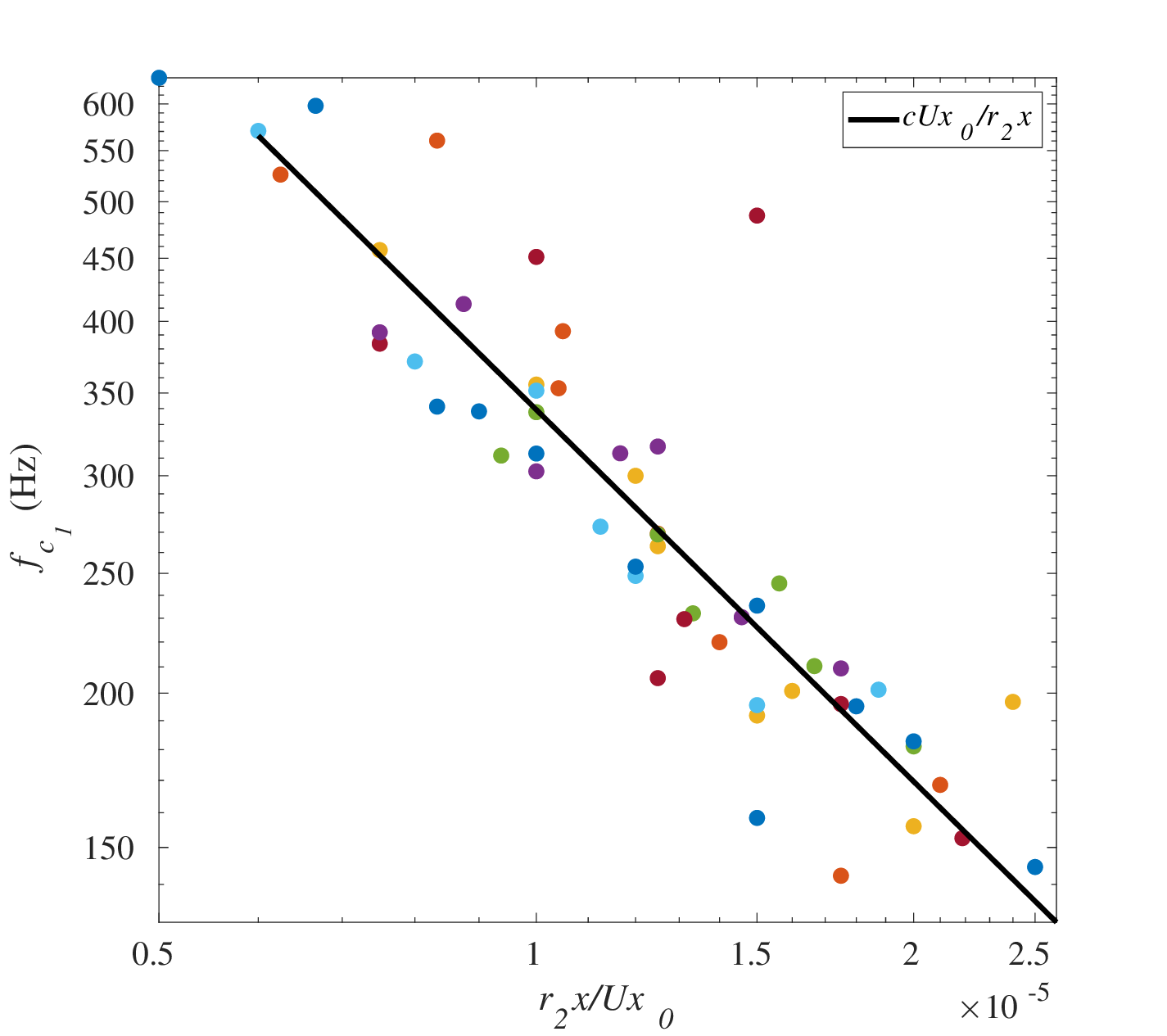}
\includegraphics[width=5cm,height=5cm]{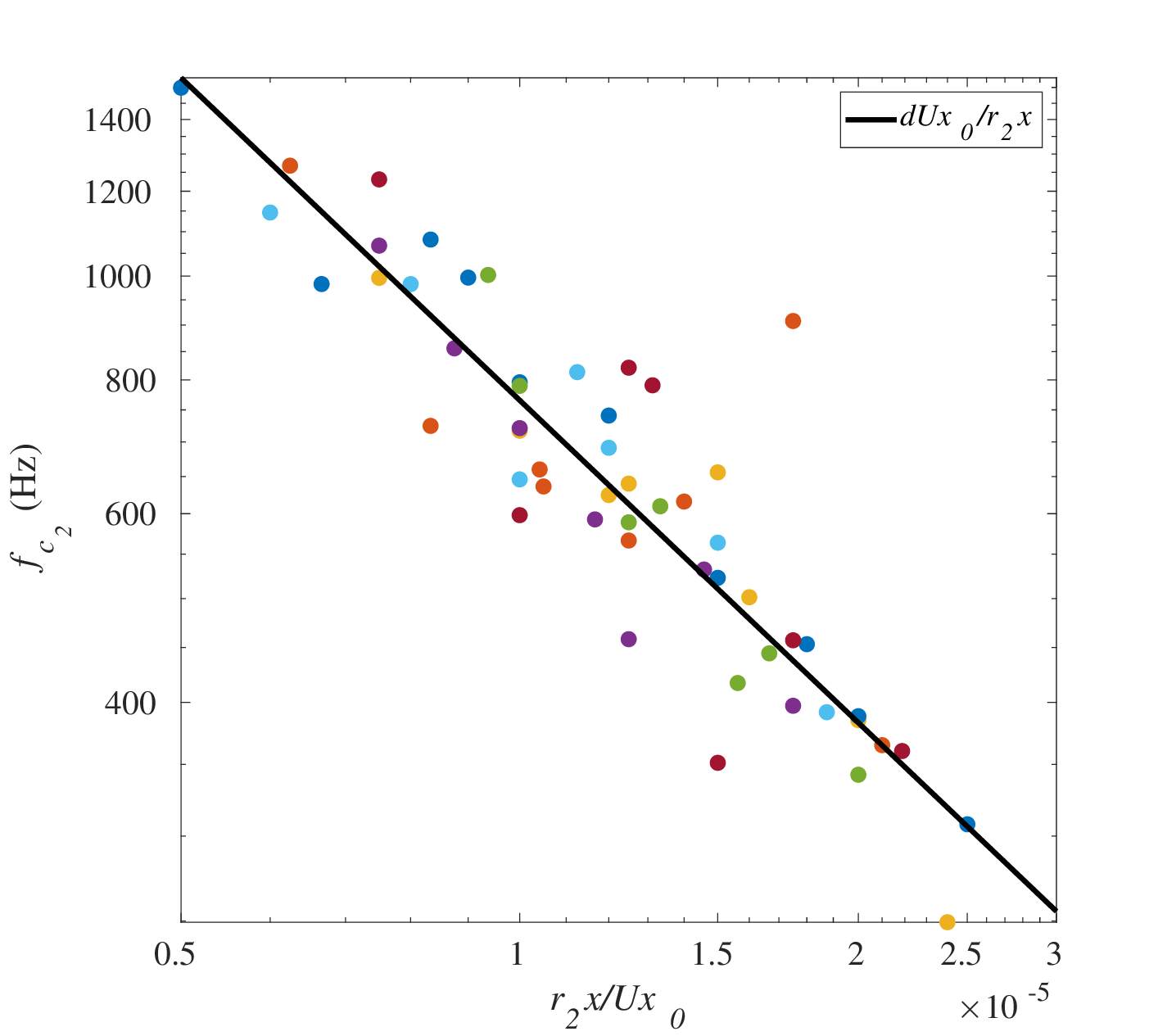}
\caption{Variation of $C_0$ (left panel), $f_{c_1}$ (middle panel), and $f_{c_2}$ (right panel) as a function of different dimensionless parameters of the system. $a=1.075$, $b=1.029\times 10^3$, $c=3.3\times 10^3$, $d=7.6\times 10^3$.} 
\label{fig3}
\end{figure*}

\subsection{Region1}
Fig. \ref{fig2}(a) shows the coherence curves of the velocity fluctuations in region 1 of the jet for different values of $r_2$ while keeping other parameters of the system constant. A typical coherence profile in this region exhibits a prominent peak (discussed in more detail in Appendix \ref{app}) at small frequencies. It is followed by a gradual decay in coherence across intermediate frequencies and eventually diminishing to the noise level for sufficiently smaller eddies with higher frequencies. Unless stated otherwise, the nature of coherence at intermediate frequencies is studied in this paper. It is observed that the coherence curve in this range of frequencies corresponds to the following equation (fit shown in Fig. \ref{fig2}(a)):

\begin{equation}
C(f) = C_0e^{-f/f_{c_1} - (f/f_{c_2})^2}
\label{eq3}   
\end{equation}
where, $C_0$ is value of coherence at $f=0$, and $f_{c_1}$ and $f_{c_2}$ are the frequencies associated with the decay of coherence and their values depend on the parameters of the system. It must be pointed out that $C_0$ in Eq. \ref{eq3} corresponds to the coherence at $f=0$ for part of the coherence curve belonging to intermediate frequencies and not the actual coherence at $f=0$. As previously mentioned, different functional form of coherence have been reported. Specifically, Davenport reported that in homogeneous turbulent low mean velocity flows $C(f) \propto $e$^{-f/f_c}$ \cite{Davenport} and in a theoretical work \cite{Tobin} $C(f) \propto $e$^{-(f/f_c)^2}$ is predicted when the flow has high mean velocity. In contrast to these observations, despite the fact that the flow has high mean velocity in the present case, we observe that the coherence is a combination of both the scenarios.

Fig.\ref{fig2}(b) shows the coherence curves for several values of different control parameters after re-scaling them by factors shown on the axes of the figure. The factor for the $x$-axis was identified by rescaling the curves independently for each parameter ($r_2$, $x$, and $U$) keeping the other two parameters constant. We must point out that we introduced an unknown length parameter ($x_0$) to make the factor on the $x$-axis of Fig.\ref{fig2}(b) dimensionless. Its value, however, is assumed constant (equal to 1) in all of the results as we actually did not control this parameter. It is evident that the coherence curves (in the intermediate frequency regime) collapse onto each other, indicating that in region 1 intermediate sized eddy structures and hence the coherence are self similar in nature. This is in contrast to a previous work \cite{Samie} in which the authors reported that coherence is not self-similar in region 1 ($r_1=0$ in their case) of a turbulent jet. Moreover, the fact that the curves collapsed by rescaling the $x$-axis of Fig.\ref{fig2}(b) with $r_2x/Ux_0$ indicates that $f_{c_1}$ and $f_{c_2}$ in Eq. \ref{eq3} are functions of $r_2x/Ux_0$. It should be noted that in the introduction section we mentioned $C(f)\propto g(r_2f/U_{rms})$ and our results indicate that $C(f)\propto g(r_2xf/Ux_0)$. This is due to the fact that for a turbulent jet flow $U_{rms}\propto U/x$ \cite{Pope}.

Least squares fit corresponding to Eq. \ref{eq3} was performed on coherence curves corresponding to all control parameters and the values of $C_0$, $f_{c_1}$, and $f_{c_2}$ were calculated. These values have been plotted in Fig. \ref{fig3}. From this it can be interpreted that 

\begin{eqnarray}
C_0&=&ae^{-b(r_2/x)^2}\\
f_{c_1}&=&cUx_0/r_2x\\
f_{c_2}&=&dUx_0/r_2x
\label{eq4}
\end{eqnarray}

\begin{figure*}[ht!]
\includegraphics[width=4cm,height=4cm]{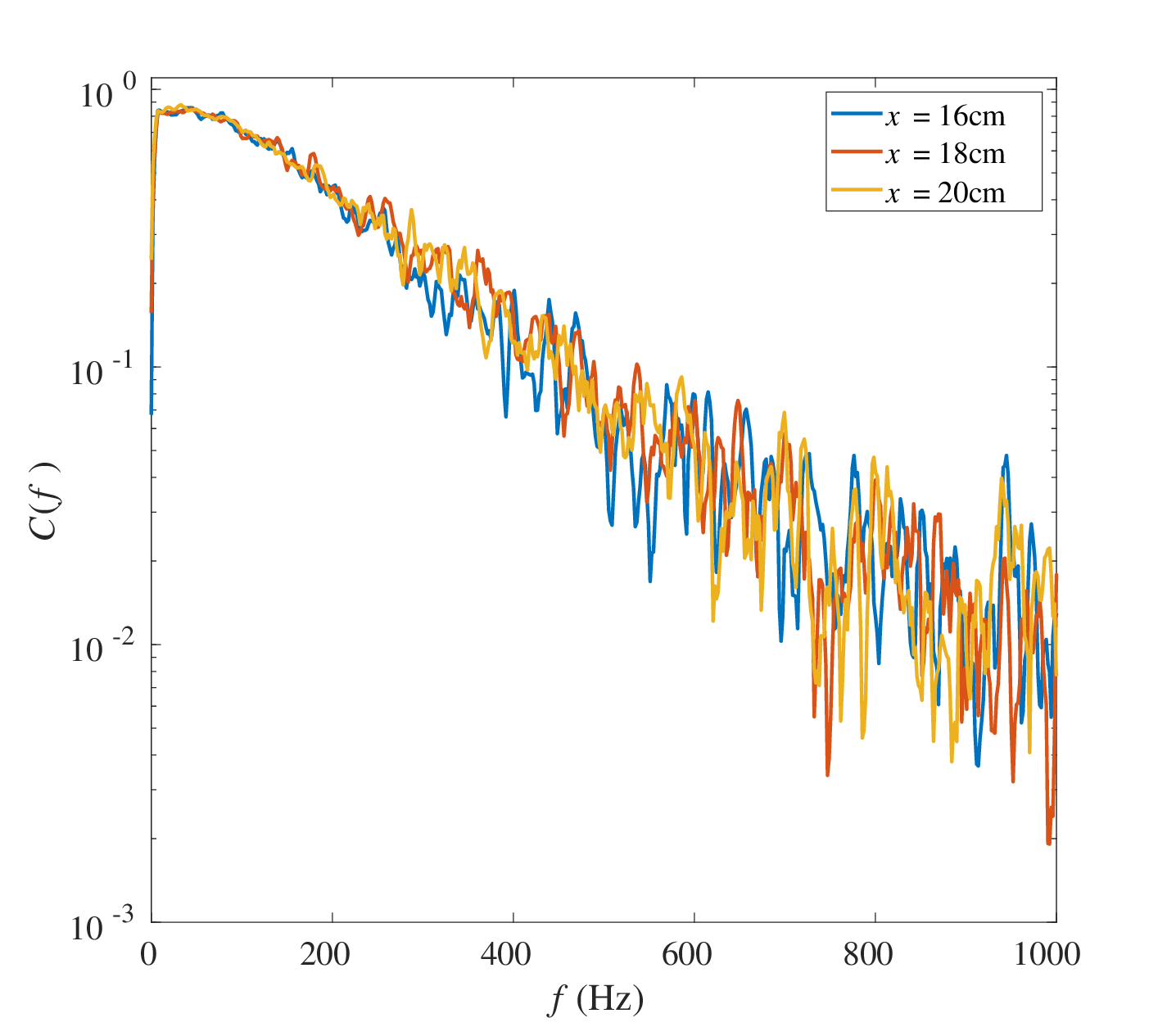}
\includegraphics[width=4cm,height=4cm]{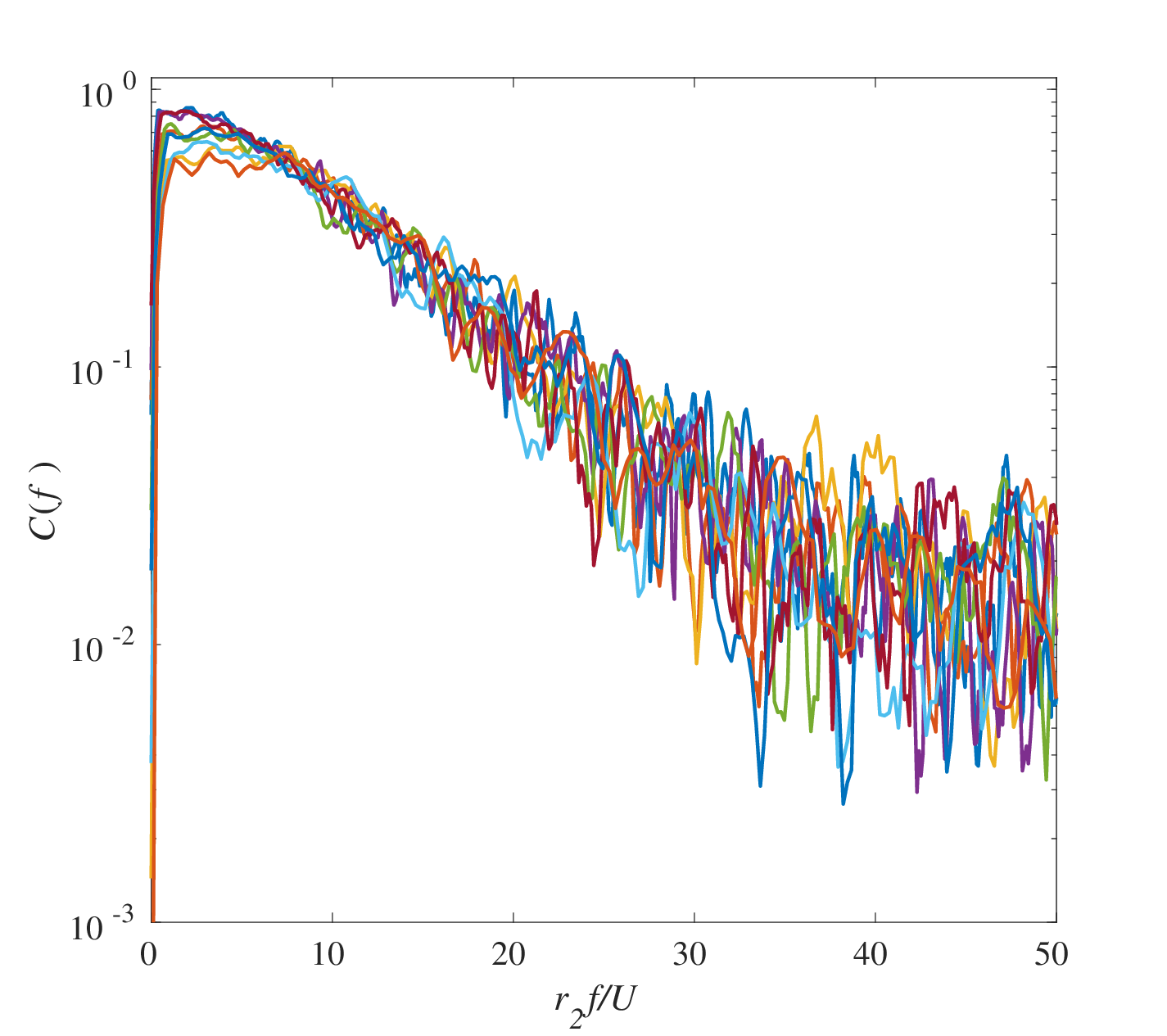}
\includegraphics[width=4cm,height=4cm]{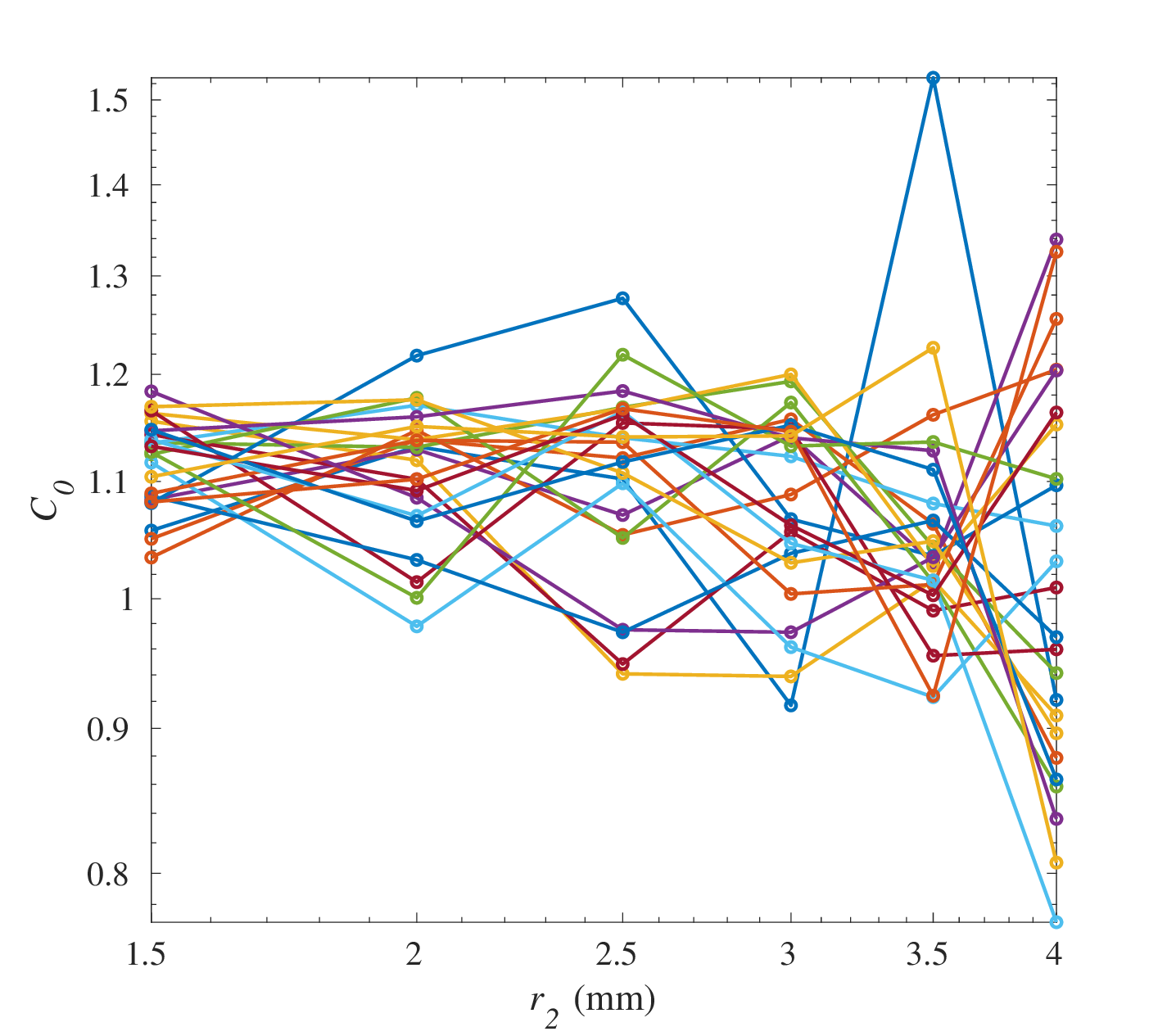}
\includegraphics[width=4cm,height=4cm]{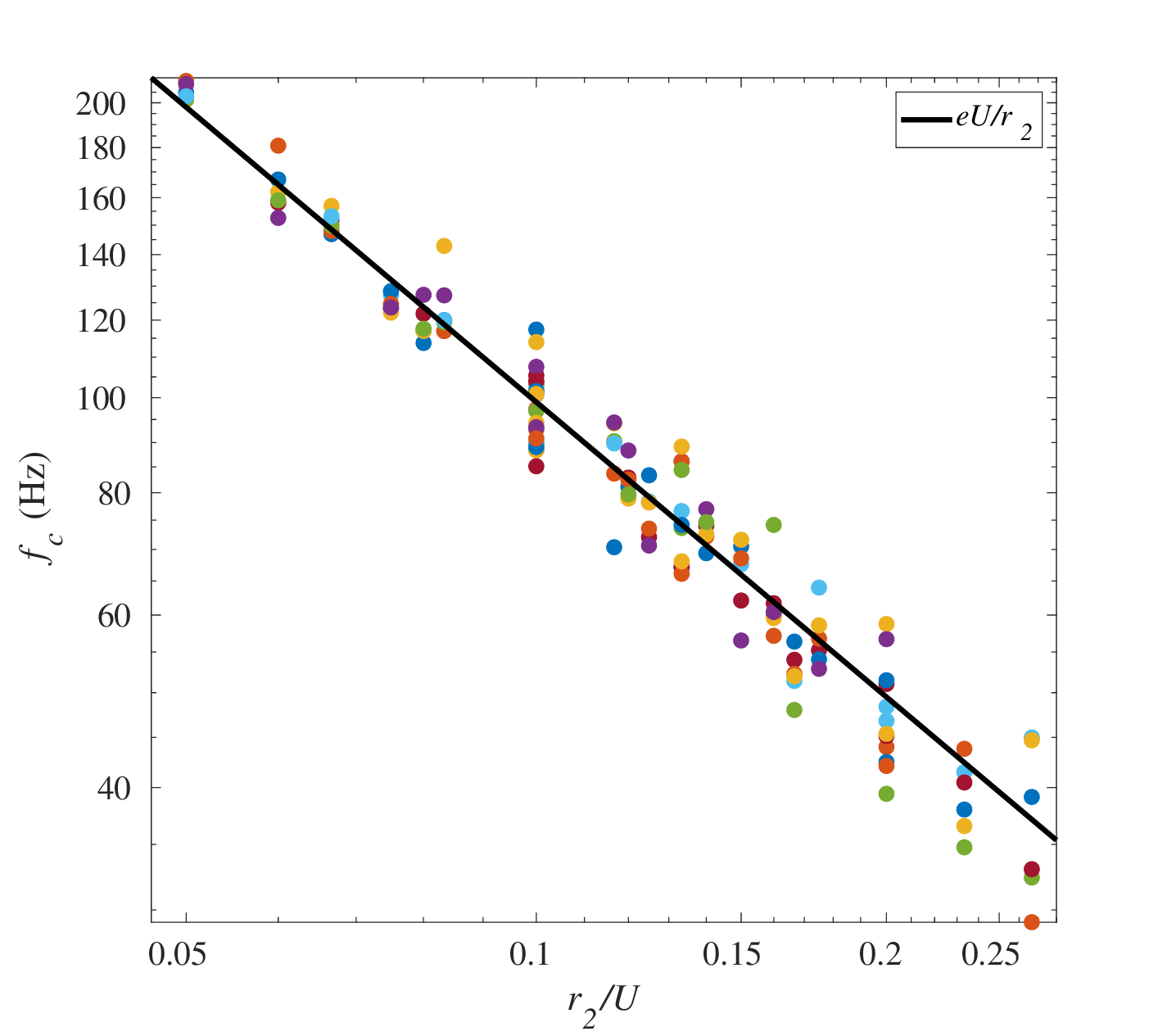}
\caption{(1st panel) Coherence curves in Region 2 of the turbulent jet as a function of $x$. (2nd panel) Re-scaled coherence curves for several parameters of the turbulent jet. (3rd panel) variation of $C_0$, and (4th panel) variation of $f_{c}$  as a function of $r_2/U$.} 
\label{fig4}
\end{figure*}

\subsection{Region2}

Fig. \ref{fig4} shows the results of coherence in region 2 where both probes lie in the region separated by the lines $0.5r_{1/2}$ and $r_{1/2}$ (Fig. \ref{setup}). As in region 1, coherence curves in this region also consist of a maxima at lower frequencies, gradual decay at intermediate frequencies, and finally coherence reaches the noise level at higher frequencies and all the results will be presented for the intermediate range. 

In the first panel, coherence curves are shown when $x$ is varied while other parameters remain constant. It must be noticed that the curves are not re-scaled on this figure and they still align with each other. This indicates that coherence does not depend on $x$ in region 2. This shows a difference between coherence in region 1 and region 2 where for the same range of $x$ and same values of $r_2$ and $U$, coherence changes in region 1 but not in 2. This could be due to the fact that the power spectra (figure in appendix) of velocities fluctuations remain almost similar for different values of $x$ in region 2.

With a least square fit on one of these curves on Fig. \ref{fig4}, it was observed that the coherence in region 2 is described by the following relation (fit not shown): 
\begin{equation}
C(f) = C_0e^{-f/f_{c}} 
\label{eq5}   
\end{equation}
This is again different from the coherence in region 1 where it has both linear and quadratic terms of frequency (Eq. \ref{eq3}). However, in the present case, coherence has only a linear term. This is due to the fact that the fluctuations in the velocity start to become similar to the mean velocity of the flow.

In the second panel of Fig. \ref{fig4}, rescaled coherence curves are shown by varying $r_2$ and $U$ and in this case coherence curves align with each other indicating that the flow is self-similar in region 2 as well.

Eq. \ref{eq5} was fitted in the intermediate frequency range of the coherence curves for different values of $r_2$ an d $U$ and the corresponding values of $C_0$ and $f_c$ were calculated; their variations are plotted on the third and fourth panels of Fig. \ref{fig4}. Results on the 3rd panel show that $C_0$ is independent of any parameter of the system and those of the 4th panel, show that $f_c$ can be fitted with $eU/r_2$ suggesting that $f_c$ is given by:
\begin{equation}
f_c=eU/r_2
\label{eq6}   
\end{equation}
It must be pointed out that $f_c$ can also be written as a function of $U_{rms}/xr_2$ as $U_{rms}\propto U/x$.

\begin{figure}[ht!]
\includegraphics[width=4cm,height=4cm]{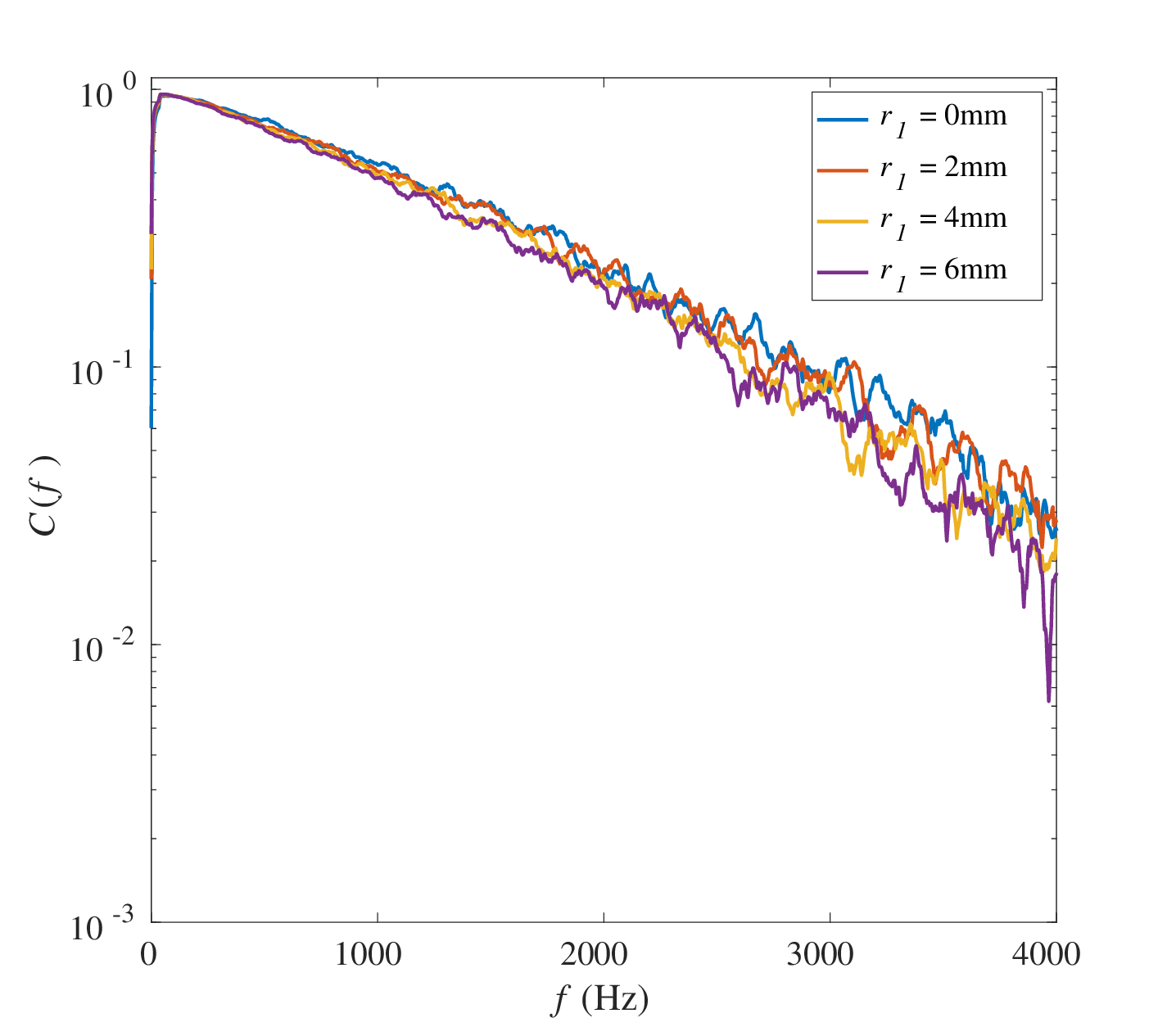}
\includegraphics[width=4cm,height=4cm]{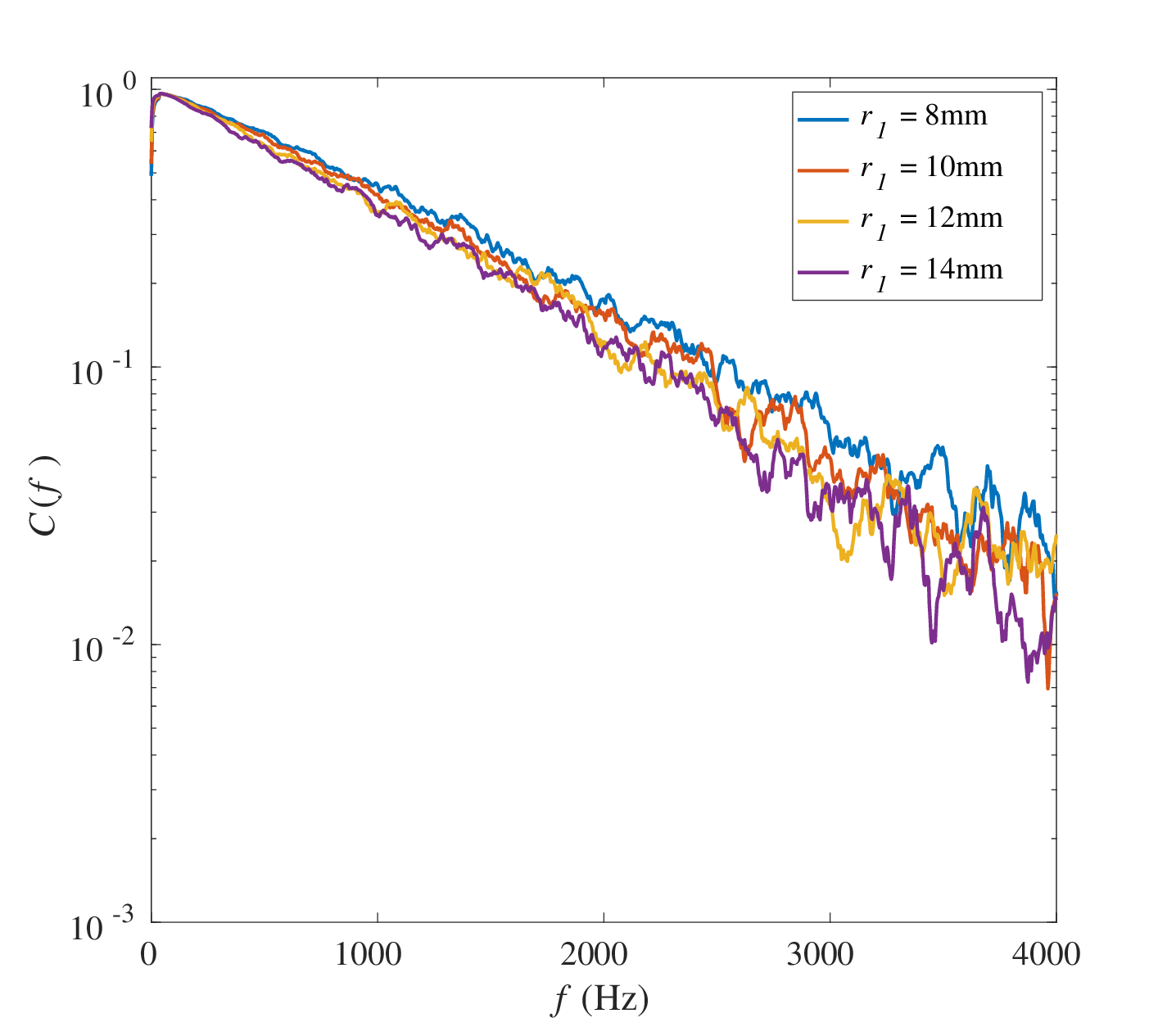}
\caption{Coherence curves as a function of $r_1$ in region 1 (left panel) and region 2 (right panel).} 
\label{fig5}
\end{figure}

\subsection{Dependence on $r_1$}
Lastly, in Fig. \ref{fig5} nature of coherence is shown as a function of $r_1$. The probes were in regions 1 and 2 in the left and right panels, respectively. It can be seen that, except for the functional form of coherence as described by Eq. \ref{eq3} and \ref{eq5}, the coherence does not change as $r_1$ is varied.

\section{Conclusions}
We measured the coherence between velocity fluctuations in two different regions in the flow of a turbulent jet. It is shown that the functional form of the coherence depends on where it is measured and in region 1 it consists of an additional term that was not theoretically predicted. Coherence is also reported to be independent of system parameters in region 2, notably, $C_0$ does not depend on any parameter and overall coherence does not depend on $x$ in this region. Other statistical properties are also observed to be invariant of $x$ in Region 2 of a turbulent jet flow. However, to the best of our knowledge, a turbulent jet is mostly studied either on the central axis or on the plane perpendicular to it. It is also observed that the coherence is self-similar in both regions. 

\section{Acknowledgments}
TS acknowledges postdoctoral funding from the CEFIPRA project 6104-1. The authors acknowledge discussions with G. Prabhudesai. 

\begin{figure}[ht!]
\includegraphics[width=5cm,height=4.5cm]{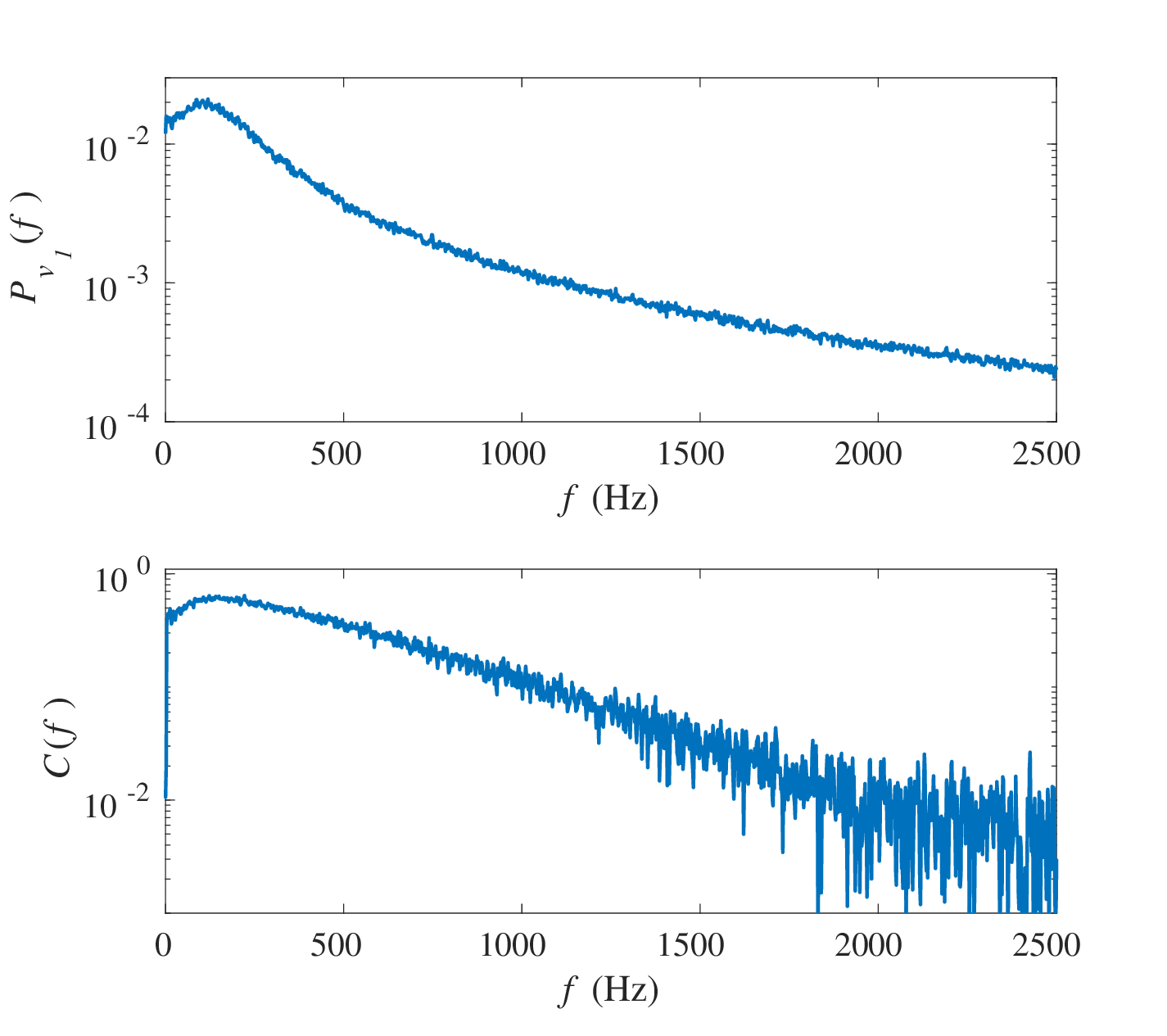}
\caption{(a) Power spectra of velocity fluctuations, and (b) corresponding coherence between velocity fluctuations.} 
\label{freq_spectra}
\end{figure}

\appendix
\section{}\label{app}

It is observed that coherence consists of a peak at lower frequencies. We also observed that the maximum of the auto and cross power spectra of velocity fluctuations, that were used to calculate the coherence, lie in the same frequency range (Fig. \ref{freq_spectra}). Although this is not a straightforward proof of the fact that there are maxima in the coherence, we believe that the maxima in the power spectra at the same frequency are responsible for this behavior in coherence. Moreover, the maxima in the power spectra correspond to the integral frequency of the flow. This implies that coherence behaves differently in the frequency range lower than the integral frequency of the flow.

\begin{figure}[ht!]
\includegraphics[width=5cm,height=5cm]{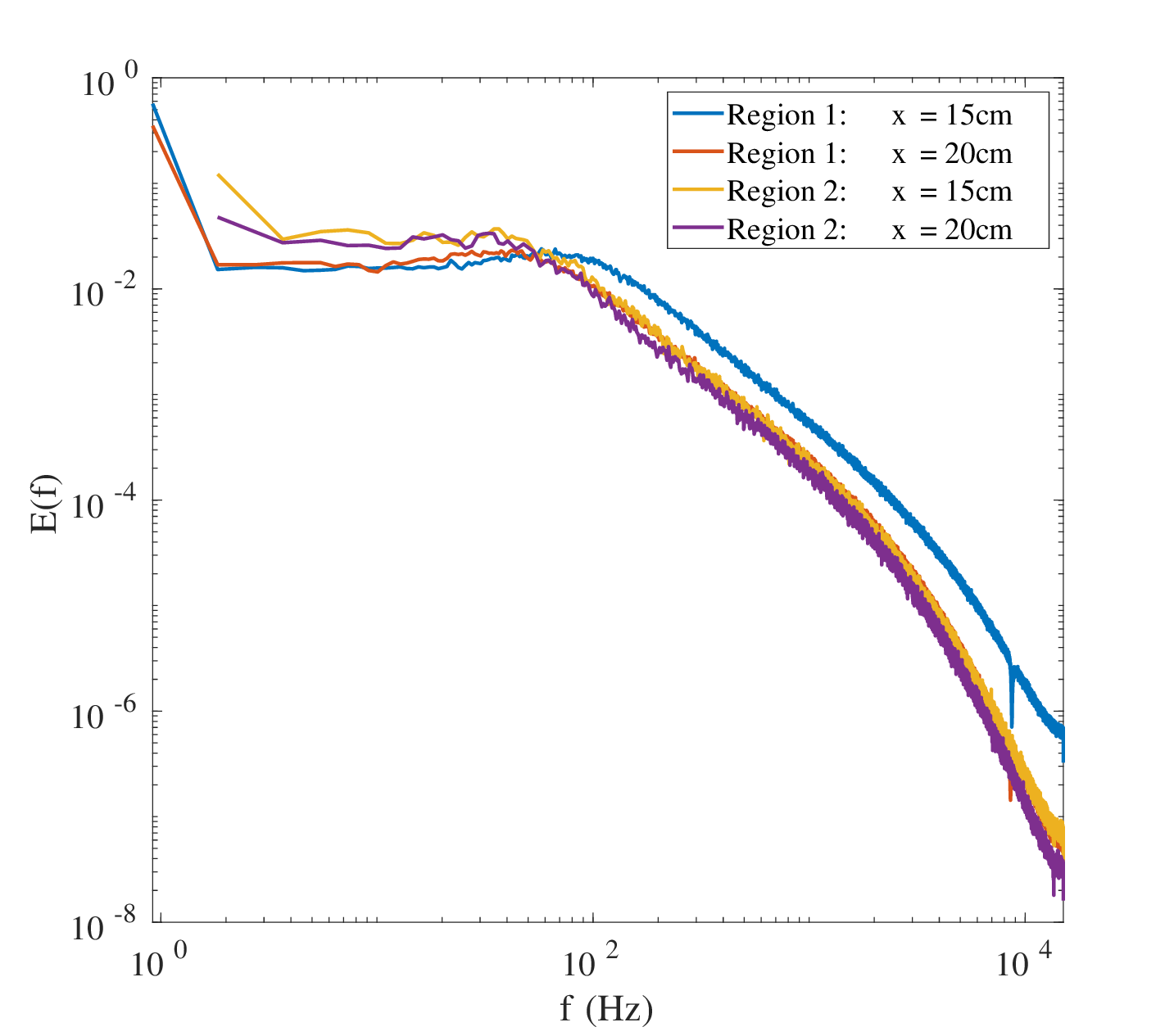}
\caption{Power spectra of velocity fluctuations for the same range of $x$ in Region 1 and 2.} 
\label{freq_spectra_app}
\end{figure}

Fig. \ref{freq_spectra_app} shows the power spectra of velocity fluctuations of the flow for two different values of $x$ in Regions 1 and 2. It can be observed that power spectra change substantially when the position of the probe changes in region 1. whereas, for the same range of $x$ in region 2, the power spectra change slightly. We believe that this is the reason why coherence is independent of $x$ in region 2 of the turbulent jet (Fig. \ref{fig4}(a)).

\bibliography{biblio}
\end{document}